\documentclass[reprint,
nofootinbib,
amsmath,amssymb,
aps
]{revtex4}

\renewcommand{\texttt}{{}}
\def\bs{\begin{subequations}}
\def\es{\end{subequations}}

\def\Cc{\mathcal{C}}

\def\Ec{\mathcal{E}}
\def\Fc{\mathcal{F}}

\def\Hc{\mathcal{H}}

\def\Lc{\mathcal{L}}

\def\Oc{\mathcal{O}}
\def\Pc{\mathcal{P}}

\def\Tc{\mathcal{T}}

\newcommand{\LF}{\left(}
\newcommand{\RF}{\right)}
\newcommand{\LT}{\left[}
\newcommand{\RT}{\right]}

\newcommand{\tia}[1]{}

\newcommand{\bea}{\begin{eqnarray}}
\newcommand{\eea}{\end{eqnarray}}
\newcommand{\beas}{\begin{eqnarray*}}
\newcommand{\eeas}{\end{eqnarray*}}
\newcommand{\bal}{\begin{aligned}}
\newcommand{\eal}{\end{aligned}}

\newcommand{\pd}{\partial}

\renewcommand{\imath}{\ensuremath{\mathrm{i}}}
\renewcommand{\vec}[1]{\ensuremath{\mathbf{#1}}}

\usepackage{dirtytalk}
\usepackage{mathtools,mathbbol}
\usepackage{hyperref,xcolor}
\usepackage{float}

\def\Hc{\mathcal{H}}
\def\Pc{\mathcal{P}}

\def\Tc{\mathcal{T}}
\def\Cc{\mathcal{C}}
\usepackage[normalem]{ulem}

\begin{document}

\title{Parity asymmetry of primordial scalar and tensor power spectra}

\author{K. Sravan Kumar$^{1}$}
\email[]{sravan.kumar@port.ac.uk}
\author{Jo\~ao Marto$^{2}$}
\email[]{jmarto@ubi.pt}
\affiliation{$^{1}$Institute of Cosmology \& Gravitation, University of Portsmouth, Dennis Sciama Building, Burnaby Road, Portsmouth, PO1 3FX, United Kingdom}
\affiliation{$^{2}$Departamento de F\'{\i}sica e Centro de Matem\'atica e Aplica\c{c}\~oes, Universidade da Beira Interior, Rua Marqu\^{e}s D'\'Avila e Bolama 6200-001 Covilh\~a, Portugal}

\date{\today}

\begin{abstract}
Although the cosmic microwave background (CMB) is largely understood to be homogeneous and isotropic, the CMB angular power spectra present anomalies that seem to break down parity symmetry at large angular scales. We argue that the primordial scalar and tensor power spectra can be parity asymmetric in our new construction of inflationary quantum fluctuations. 
 Our formulation stems from the foundational questions of quantum field theory in curved spacetime in which we impose geometric superselection rules to the vacuum structure for (single-field) inflationary quantum fluctuations based on discrete spacetime transformations ($\Pc\Tc$). As a result, we estimate the amplitude of power asymmetry in the scalar and tensor sectors at different scales of $ 10^{-4} {\rm Mpc^{-1}}\lesssim k\lesssim  10^{-3}{\rm Mpc^{-1}}$. 
In particular, we predict the parity asymmetry for the primordial gravitational waves (PGWs) and quantify it for different models, like Starobinsky and $\alpha-$attractor single-field inflationary scenarios.
\end{abstract}


 \maketitle

\section{Introduction}
\label{introduction}

The inflationary paradigm in the early Universe lays a strong foundation for the successful description of the CMB and the large-scale structure of the Universe that is consistent with the observation of scalar spectral index $n_s=0.9649\pm 0.0042$ for $50-60$ e-foldings of quasi-de Sitter expansion \cite{Starobinsky:1980te,Starobinsky:1979ty,Planck:2018jri}. Besides its apparent success, some CMB anomalies seem to arise and challenge any current model of inflation. 
Concretely, CMB parity anomalies appear at large angular scales \cite{Planck:2019evm}, seen from NASA's Wilkinson Microwave Anisotropy Probe to the latest Planck data. 
It was recently shown that most CMB anomalies originated from prevalent parity odd preferred character of CMB \cite{Gaztanaga:2024vtr}. 
 In short, these anomalies are explicit in the two-point temperature correlations at the parity conjugate points of the CMB sky \cite{Gaztanaga:2024vtr}. This study has shown that the CMB is statistically homogeneous and isotropic but parity asymmetric. 
 This is a significant challenge to the standard (inflationary) cosmology model, which does not explain their origin.  
 These anomalies appear significantly at low-multipoles $\ell \lesssim 30$ \cite{Schwarz:2015cma}. 
{To better characterize the parity asymmetry in the two-point correlations, we define the fractional difference in the power spectra at }$\textbf{x}$ and $-\textbf{x}$
\begin{equation}
\begin{aligned}
	A(k) & = \frac{\Pc_{\zeta+}\LF k,\,\hat{\textbf{x}} \RF - \Pc_{\zeta-}\LF k,\,-\hat{\textbf{x}} \RF }{4\Pc_{\zeta}} \\
 T(k) & = \frac{\Pc_{h+}\LF k,\,\hat{\textbf{x}} \RF - \Pc_{h-}\LF k,\,-\hat{\textbf{x}} \RF }{4\Pc_{h}}
 \end{aligned}
	\label{Akg}
\end{equation}
where $\zeta,\,h$ is the curvature perturbation and the tensor fluctuation.  The subscripts with $\pm$ indicate the respected quantities evaluated at parity conjugate points. The power spectra in the denominator are the usual approximately scale-invariant ones
\begin{equation}
    \Pc_{\zeta} \approx A_s\LF \frac{k}{k_\ast} \RF^{n_s-1},\quad  \Pc_{h} \approx A_t\LF \frac{k}{k_\ast} \RF^{n_t}\,,
\end{equation}
where $A_s= 2.2\times 10^{-9}$ is the scalar amplitude measured by Planck \cite{Planck:2018jri} and the amplitude of tensor power spectra is bounded by $r= \frac{A_t}{A_s}<0.036$. The scale $k_\ast = 0.05\,{\rm Mpc}^{-1}$ which corresponds to angular scales of $\lesssim 1^\circ$ in the CMB sky. The quantities $n_s,\, n_t$ are called scalar and tensor spectral index, respectively.    

A quantum mechanical explanation of how one can generate the quantities \eqref{Akg} from single-field slow-inflationary quantum fluctuations forms the crux of this paper. {The parity asymmetry measured by the quantities \eqref{Akg} would result in oscillation in the angular power spectra of even-odd multipoles.} The implications of $A(k)$ derived in the context of this study are tested with observations in \cite{Gaztanaga:2024vtr}, and here we derive predictions (for the first time) for the parity asymmetry of primordial gravitational waves for the large-scale $\ell\lesssim 30$.

In this paper, we show inflationary quantum fluctuations generate parity asymmetry in both CMB and primordial gravitational waves on large scales. Our framework is focused on an intricate understanding of how the spontaneous breaking of {time reversal symmetry} during inflation can generate observational signals in the form of parity asymmetric primordial power spectra. 
Thus, we only focus on single-field slow-roll inflation and propose a vacuum for quantum fluctuations, leading to new observational effects. We direct the reader to \cite{Kumar:2023ctp,Kumar:2023hbj,Kumar:2024oxf,GKM} for the formal aspects of the new framework of quantum field theory in curved spacetime that we implemented here in the context of inflationary quantum fluctuations. 

\section{Inflationary power spectra and the role of $\Pc\Tc$ transformations}
\label{sec:questions}

By quantizing inflationary fluctuations, we actually deal with quantum field theory (QFT) in curved spacetime. It is worth recalling that QFT in Minkowski spacetime ($ds^2= -dt_p^2+d\textbf{x}^2$) can be seen as the unification of special relativity and quantum mechanics achieved by demanding the field operators to commute for space-like distances and the time ($t_p$) is being treated as a parameter \cite{Donoghue:2017pgk,Donoghue:2019ecz} (See also Appendix.~\ref{sec:DQM} for further details).  

In the case of inflationary cosmology, we usually quantize both gravitational and matter degrees of freedom \cite{Martin:2004um,Baumann:2018muz,Kinney:2009vz} taking a classical notion of time. However, the concept of time in quantum theory is very different from the classical one \cite{Rovelli:2004tv}.
Since we quantize gravitational degrees of freedom, we can interpret inflationary quantum fluctuations as features of (linearized) quantum gravity \cite{Martin:2004um}. Note that, in the canonical quantum gravity that emerges through the Wheeler-de Witt equation, time does not appear explicitly, which indicates a difficulty in defining positive/negative frequencies, like we usually do following the standard Schr\"{o}dinger equation. This is famously known as the problem of time in quantum cosmology \cite{Kiefer:2007ria,Rovelli:2004tv}.
Quantum theory is always time-symmetric. The arrow of time emerges only after we specify initial and final states \cite{Hartle:2013tm}. This implies that we could formulate QFT in curved spacetime in a time-symmetric way and then impose initial conditions. 

Understanding quantum fields in de Sitter (dS) spacetime is crucial for the study of inflationary quantum fluctuations since inflationary spacetime is quasi-dS (qdS) \cite{Starobinsky:1980te}. The dS spacetime, which in flat Friedman-Lema\^itre-Robertson-Walker (FLRW) coordinates looks like
\begin{equation}
	\begin{aligned}
	ds^2 & = -dt^2 + a(t)^2d\textbf{x}^2
	            & = \frac{1}{H^2\tau^2}\LF -d\tau^2+ d\textbf{x}^2  \RF\,. 
	\end{aligned}
	\label{dsmetric}
\end{equation}
where $d\tau= \frac{dt}{a}$ is the conformal time  and the
scale factor is given by
\begin{equation}
	a(t) = e^{Ht},\quad H^2 = \LF\frac{1}{a}\frac{da}{dt}\RF^2={\rm const}\,.
\end{equation}
Note that dS metric \eqref{dsmetric} is 
\begin{equation}
	\Pc\Tc: \tau\to -\tau,\, \textbf{x}\to -\textbf{x} 
\end{equation}
symmetric. 
The scalar curvature of dS spacetime is $R=12H^2$, which does not tell whether the Hubble parameter is positive or negative. Each point in dS is surrounded by a comoving horizon given by the radius
\begin{equation}
	r_H = \Big\vert  \frac{1}{aH} \Big\vert
	\label{horizonds}
\end{equation}
One very simple observation we can make from \eqref{dsmetric} is that  
\begin{equation}
	{\rm Expanding\,Universe:} \implies \Bigg\{ \begin{matrix}
		\begin{aligned}
		t: -\infty \to \infty\quad & H>0\\ 
		t:	\infty \to -\infty\quad & H<0 \,.
			\end{aligned}
	\end{matrix}
	\label{Expanding-U-eq}
\end{equation}
The arrow of time corresponding to the expanding Universe can be designated as $\tau:  \mp \infty \to 0$ (with $H\to -H$ and $t\to -t$). To describe the expanding Universe, we must think of the scale factor as a clock, which means
we identify the expansion of the Universe with the shrinking size of the comoving horizon. From \eqref{Expanding-U-eq}, we learn that we have two arrows of time to describe an expanding Universe. In the literature, it is usually considered that $\tau<0$ ($H>0$, $t: -\infty \to \infty $) to describe an expanding universe \cite{Birrell:1982ix,Allen:1985ux}, but this would leave the existence of another Universe with the opposite arrow of time $\tau>0$ (which is still expanding for $H<0$ with $t: \infty \to -\infty$). Most often, the studies treat these Universes as entangled ones with unitarity being lost for an observer living in either of the Universe.  Schr\"{o}dinger in 1956 \cite{Schrodinger1956,Parikh_2003} has demanded to describe one Universe with two arrows of time (quantum mechanically). This presents a new understanding of quantum theory. In Appendix.~\ref{sec:DQM}, we present the appearance of two arrows of time in quantum mechanics and review a reconstruction of it with what we call direct-sum quantum theory, which is developed in the previous works \cite{Kumar:2023ctp,Kumar:2023hbj,Kumar:2024oxf,GKM}. According to this construction, a quantum state is expressed as direct-sum of two components in the Hilbert spaces with opposite arrows of time that are attached to the parity conjugate regions of physical space. Such Hilbert spaces are called the geometric superselection sectors because (component) states in these do not form a superposition but are joined by a direct-sum operation in the total (direct-sum) Hilbert space. The framework is extended to rethink QFT in Minkowski spacetime with two arrows of time, where a single quantum field operator is written as a direct-sum of two components in a direct-sum Fock space with geometric superselection sectors (See Appendix.~\ref{sec:DQM}). {This construction is what we call direct-sum quantum field theory (DQFT). In direct-sum quantum theory, we attached the Hilbert and Fock spaces with the regions of physical space related by discrete transformations, which are geometric superselection rules. } In Appendix.~\ref{app:dS}, we present the direct-sum quantization of a Klein-Gordon field in dS spacetime that takes into account the two arrows of time \eqref{Expanding-U-eq}.  

{Inflationary spacetime breaks the time-reversal symmetry of the dS spacetime \eqref{Expanding-U-eq}.
In our approach, an inflationary quantum fluctuation is represented by a direct-sum of two {components} generated in a direct-sum vacuum based on $\Pc\Tc$ transformations in the gravitational context. This means within the co-moving horizon radius $r_H\sim \vert \frac{1}{aH}\vert $ a quantum fluctuation evolves forward in time in the spatial region spanned by the angular coordinates $\LF \theta,\,\varphi \RF$ and evolves backward in time at the angular coordinates $\LF \pi-\theta,\,\pi+\varphi \RF$. 
As inflation proceeds, when a quantum fluctuation exits the horizon on the two antipodal points, it is not the same due to the time asymmetry created by the inflationary expansion.  This creates parity asymmetry in the CMB sky and the primordial gravitational wave background. }
{Therefore, in our framework of quantization, we can explicitly see how power asymmetry in the primordial correlations is a signature of the spontaneous breaking of time-reversal symmetry in the expanding Universe.}

 We now define the total Fock space vacuum in dS as the direct-sum of two vacua
\begin{equation}
	\vert 0\rangle_{dS} = \vert 0\rangle_{dS+}\oplus \vert 0\rangle_{dS-}  = \begin{pmatrix}
		\vert 0\rangle_{dS+} \\ 
		\vert 0 \rangle_{dS-}
	\end{pmatrix}\,. 
\label{dSvacmat}
\end{equation}
{A quantum field now in the vacuum $\vert 0\rangle_{dS} $ gets created everywhere as a direct-sum of the two {components} in the
vacuum $\vert 0\rangle_{dS+}$ at the position $ \textbf{x}$ and  vacuum $\vert 0\rangle_{dS-}$ at the position $- \textbf{x}$.}
{In the case of the quantization of a massless scalar field in dS spacetime, the two-point correlations in the $\Pc\Tc$ related vacuums $\vert 0\rangle_{+}$ and $\vert 0\rangle_{-}$ are equal (See Appendix.~\ref{app:dS}). Further understanding of this formulation of QFT in dS can be found in \cite{Kumar:2023ctp,Kumar:2024oxf}. }  

 Unlike dS spacetime, the inflationary spacetime (which is quasi de Sitter) does not have $\Pc\Tc$-symmetry, therefore naturally one would expect the $\Pc\Tc$-symmetry to be spontaneously broken at the quantum level. In the context of inflation, we quantize
 metric and matter degrees of freedom to find an
 effective quantum correction to the classical spacetime.
Assuming that inflationary quantum fluctuations {described as a direct-sum based on the} $\Pc\Tc$ transformations, 
we write the canonical field operator as a direct-sum {which means \cite{Conway}}
\begin{equation}
	\begin{aligned}
	\hat{v} & = \frac{1}{\sqrt{2}} \hat{v}_{+}\LF \tau,\, \textbf{x} \RF \oplus  \frac{1}{\sqrt{2}} \hat{v}_{-}\LF -\tau,\, -\textbf{x} \RF \\ & = \frac{1}{\sqrt{2}} \begin{pmatrix}
		\hat{v}_+ \LF \tau,\,\textbf{x} \RF & 0 \\ 
		0 & \hat{v}_- \LF -\tau,\, -\textbf{x} \RF
	\end{pmatrix}
\end{aligned}
\label{fieldmat}
\end{equation}
where $\hat{v}$ is the quantum counterpart of the field redefinition (which is Mukhanov-Sasaki variable)  $v=a\zeta \dot{\phi}/H$ of the curvature perturbation ($\zeta$) \cite{Baumann:2009ds,Kinney:2009vz} where $H$ is the Hubble parameter during inflation, $\phi$ is the inflaton field and overdot represents derivative with respect to cosmic time ($t$).
The total vacuum in the quasi de Sitter (qdS) spacetime is a direct-sum given by 
\begin{equation}
\vert 0\rangle_{\rm qdS} = \vert 0 \rangle_{\rm qdS_{I}} \oplus \vert 0\rangle_{\rm qdS_{II}} = \begin{pmatrix}
	\vert 0\rangle_{\rm qdS_{I}} \\ 
	\vert 0\rangle_{\rm qdS_{II}}
\end{pmatrix}
\label{qdSmat}
\end{equation}
The quantum field operators acting on the vacua $\hat{v}_{+}\LF \tau,\, \textbf{x} \RF \vert 0\rangle_{\rm qdS_{II}} $ and $\hat{v}_{-}\LF -\tau,\, -\textbf{x} \RF\vert 0\rangle_{\rm qdS_{II}}$ leads to the description of a quantum fluctuation evolving forward in time ($\tau<0,\, H>0,\, t: -\infty\to \infty$) at  $\textbf{x}$ and evolving backward in time ($\tau>0,\, H<0,\, t: \infty\to -\infty$) at $ -\textbf{x} $.

Expanding the fields in terms of creation and annihilation operators $c_{\LF\pm \RF\textbf{k}},\, c_{\LF\pm \RF\textbf{k}}^\dagger$, 
\begin{equation}
	\begin{aligned}
& \hat{v}_{\pm }=   
\int \frac{ d^3k}{\LF 2\pi \RF^{3/2}} \Bigg[ c_{\LF\pm \RF\textbf{k}} {v}_{\pm,\,k} e^{\pm i\textbf{k}\cdot \textbf{x}} + c_{\LF\pm \RF\textbf{k}}^\dagger {v}_{\pm,\,k}^\ast e^{\mp i\textbf{k}\cdot \textbf{x}} \Bigg]
\end{aligned}
\label{vid}
\end{equation}
we define the
qdS vacua as
\begin{equation}
	c_{\LF + \RF\textbf{k}}\vert 0\rangle_{\rm qdS_I} = 0\quad c_{\LF - \RF\textbf{k}}\vert 0\rangle_{\rm qdS_{II}} = 0
	\label{qds1}
\end{equation}
and $ {v}_{\pm,\,k}$ are the mode function obtained by solving Mukhanov-Sasaki (MS) equation for $v_{\pm}$ 
\begin{equation}
	v_{\pm,\,k}^{\prime\prime}+ \LF k^2-\frac{{\nu}_s^{\LF \pm\RF 2}-\frac{1}{4}}{\tau^2} \RF v_{\pm,\,k}^2 =0\,.
	\label{MS-equation}
\end{equation}
where
\begin{equation}
	\nu_s^{\pm} \approx \frac{3}{2}\pm\epsilon\pm\frac{\eta}{2}
\end{equation}
We note that $
\big[c_{\LF + \RF\textbf{k}},\, c_{\LF - \RF\textbf{k}^\prime}\big] = 0,\quad \big[c^\dagger_{\LF + \RF\textbf{k}},\, c^\dagger_{\LF - \RF\textbf{k}^\prime}\big]  = 0 $ which means the modes $v_{\pm,\,k}$ do not have causal connection.
Notice that, unlike the dS case (Appendix.~\ref{app:dS}), MS-equation \eqref{MS-equation} is not symmetric under time reversal because there is an additional time dependence which enters through the nearly constant variable $\nu_s$ which contains the slow-roll parameters $\LF \epsilon = -\dot{H}/H^2,\:\: \eta = \dot{\epsilon}/(H\epsilon)\RF$. 


\begin{equation}
	\begin{aligned}
&	{v}_{\pm ,\,k}   = \frac{\sqrt{\mp \pi \tau}}{2} e^{\LF i\nu_s^{\pm}+1\RF} \Bigg[C_k^{\pm} H^{(1)}_{\nu_s^{\pm}}\LF \mp k \tau \RF
+ D_k^{\pm} H^{(2)}_{\nu_s^{\pm}}\LF \mp k  \tau \RF\Bigg].
	\label{new-vac1}
	\end{aligned}
\end{equation}
The above mode functions correspond to the creation of positive frequency modes in the limit $\tau\to \mp\infty$ for the case $\LF C_k^{\pm},\, D_k^{\pm} \RF = \LF 1,\,0 \RF $ which corresponds to the standard BD state that satisfies the Wronskian $v_{\pm,k}v_{\pm,k}^{\prime\ast}-v_{\pm,k}^\ast v_{\pm,k}^{\prime}=\pm \:i  \:\:(\implies \vert C_k^{\pm}\vert^2-\vert D_k^{\pm}\vert^2=1)$. 
Notice that the Wronskian condition equating to $-i$ corresponds to the canonical commutation relation for a reversed arrow of time \cite{Donoghue:2019ecz}
\begin{equation}
	\Big[ \hat{v}_{-}\LF -\tau,\,-\textbf{x} \RF,\,  \hat{\Pi}_{-}\LF -\tau,\,-\textbf{x}^\prime \RF \Big] = -i \delta\LF \textbf{x}-\textbf{x}^\prime \RF\,. 
\end{equation}
while the canonical commutation relation for $\hat v_+$ field is 
\begin{equation}
	\Big[ \hat{v}_{+}\LF \tau,\,\textbf{x} \RF,\,  \hat{\Pi}_{+}\LF \tau,\,\textbf{x}^\prime \RF \Big] = i \delta\LF \textbf{x}-\textbf{x}^\prime \RF\,. 
\end{equation}
Expanding \eqref{new-vac1} up to the leading order in $\LF \epsilon, \eta \RF$ we get 
\begin{equation}
	\begin{aligned}
	{v}_{\pm,\,k} &  \approx \sqrt{\frac{1}{2k}} e^{\mp ik\tau}\LF 1\mp\frac{i}{k\tau} \RF \\ & 
	\pm \LF \epsilon+\frac{\eta}{2} \RF \frac{\sqrt{\pi}}{2\sqrt{k}} \sqrt{\mp k\tau} \frac{\pd H^{(1)}_{\nu_s^{\pm}}\LF \mp k\tau\RF}{\pd \nu_s^{\pm}}\Big\vert_{\nu_s^{\pm}=3/2}
\label{new-vac1BD}
\end{aligned}
\end{equation}
Comparing the two-mode functions in \eqref{new-vac1BD}, we can deduce that they get different slow-roll (quantum) corrections. 
The fact that MS-equation has a time dependence in terms of $\epsilon,\, \eta$ is crucial to probe the nature of quantum fluctuations. These quantities change sign under discrete spacetime transformation and this has to be understood in a completely quantum mechanical sense. Our approach implies a {quantum fluctuation (a single degree of freedom) as a direct-sum of a {component} that } propagates with respect to an arrow of time ($\tau: -\infty \to 0$) at spatial position $\textbf{x}$  {another {component}} that propagates with respect to the opposite arrow of time ($\tau: \infty \to 0$) at the spatial position $-\textbf{x}$.  {According to this formulation, when a fluctuation exits the horizon, a {component} of that, i.e., $\hat{v}_+ \vert 0\rangle_+$ which exits the horizon on one side and another {component} of that $\hat{v}_+ \vert 0\rangle_-$ exists the horizon on the other side. }

Notice that the time reversal transformations are 
\begin{equation}
	t\to -t \implies H\to -H,\quad \epsilon\to -\epsilon,\quad \eta\to -\eta\,. 
	\label{timerevqds}
\end{equation}
which can be understood in the following sense. First, we need to formulate the meaning of the time reversal operation. Logically, if the fluctuation propagates forward in time in a slow-roll background, the fluctuation that goes backward in time experiences spacetime as a "slow-climb"\footnote{In the standard slow-roll we have scalar field rolling down the potential whose time reversal can be understood as a phantom field that is slowly climbing the potential \cite{Piao:2004tq}. } which is given by reversing the signs of the parameters $\LF \epsilon,\, \eta \RF$ as stated in \eqref{timerevqds}. We impose this time reversal operation in a completely quantum mechanical sense, and this has no classical meaning, i.e., our background (classical) dynamics is completely determined by Friedmann equations, and we do not apply at all time reversal to the classical background. Since we treat time differently at quantum level we restrain ourselves from any intuition from classical physics.\footnote{The notion of time is a non-trivial concept in physics and its meaning varies in different contexts. We suggest the reader \cite{Rovelli:2004tv} for an extended physical discussion. Our statement that a quantum state evolving backward in time has no classical analog is deeply rooted in the quantum gravity concept of time as it is presented on p. 184 of \cite{Rovelli:2004tv}.} Since dynamics of quantum fields emerge from MS-equation, the functions $\LF \epsilon,\, \eta \RF$ are now treated along with time as parameters to specify the nature of quantum states. This would encode a subtle difference between quantum fluctuations propagating forward and backward in time {at the spatial positions divided by parity}.

\section{Inflationary power spectra and consequences of parity asymmetry}
\label{section3}

 As we learned in the previous section, {inflationary quantum fluctuations now behave differently when exiting the horizon at the antipodal points. }The two-point correlations of these fluctuations can be computed as
\begin{equation}
	\begin{aligned}
			{}_{\rm qdS_{I}}\langle 0 \vert  \hat{v}_{+} \hat{v}_{+} \vert 0\rangle_{\rm qdS_{I}}  & =  \frac{4\pi}{\LF 2\pi \RF^3} \int \frac{dk}{k} \frac{\sin k \xi}{k\xi} k^3 \vert v_{+,\,k}\vert^2 \\ 
				{}_{\rm qdS_{II}}\langle 0 \vert  \hat{v}_{-} \hat{v}_{-} \vert 0\rangle_{\rm qdS_{II}} &  =  \frac{4\pi}{\LF 2\pi \RF^3} \int \frac{dk}{k} \frac{\sin k \xi}{k\xi} k^3 \vert v_{-,\,k}\vert^2\,
	\end{aligned}
\label{power-spectrav}
\end{equation}
Substituting fields \eqref{new-vac1BD} in the above expressions turn the two correlations different. In the case of exact dS (discussed in Appendix.~\ref{app:dS}), we obtain equal correlations, but in the case of qdS, there is a difference because the background spacetime is not time-reversal symmetric. 
As we know, curvature perturbation is frozen on super-horizon scales. 
To calculate the two-point correlations of curvature perturbation on super-horizon scales, we re-scale the canonical fields with the classical background quantities as
\begin{equation}
	\begin{aligned}
{}_{\rm qdS}\langle 0 \vert \zeta_\textbf{k} \zeta_{\textbf{k}^\prime} \vert 0\rangle_{\rm qdS} & =  \LF \frac{1}{2a^2\epsilon}\RF\Bigg\vert_{\rm clas.} 
\frac{1}{2}\Big[{}_{\rm qdS_I}\langle 0 \vert \hat{v}_{+\,\textbf{k}} \hat{v}_{+\,\textbf{k}^\prime} \vert 0\rangle_{\rm qdS_I}\\ & \quad +{}_{\rm qdS_{II}}\langle 0 \vert \hat{v}_{-\,\textbf{k}} \hat{v}_{-\,\textbf{k}^\prime} \vert 0\rangle_{\rm qdS_{II}}\Big] \\ 
& = \frac{2\pi^2}{k^3}\LF P_{\zeta_{+}}+P_{\zeta_{-}} \RF \delta\LF \textbf{k}+\textbf{k}^\prime \RF\,, 
\end{aligned}
\label{twocor}
\end{equation}
{In deriving \eqref{twocor} we must use \eqref{qdSmat} and \eqref{fieldmat} which implies the direct-sum field operator $\hat{v}$  acting on the vacuum $\vert 0 \rangle_{\rm qdS}$ gives
\begin{equation}
\hat{v}\vert 0 \rangle_{\rm qdS} = \frac{1}{\sqrt{2}}	
\biggl( \hat{v}_+ \oplus \hat{v}_- \biggr) \;\; 
\biggl(\vert 0 \rangle_{\rm qdS_I} \oplus \vert 0\rangle_{\rm qdS_{II}} \biggr)
= \frac{1}{\sqrt{2}}\begin{pmatrix}
\hat{v}_+ \vert 0\rangle_{\rm qdS_{I}} \\ 
\hat{v}_{-} \vert 0 \rangle_{\rm qdS_{II}}
\end{pmatrix}
\end{equation} 
 With appropriate normalization, we evaluate the power spectrum at the parity conjugate points of horizon exit as
\begin{equation}
	\begin{aligned}
	P_{\zeta_{\pm}}&  = \frac{k^3}{2\pi^2}\frac{1}{2a^2\epsilon} \vert v_{\pm,\,k}\vert^2\Bigg\vert_{\tau = \pm \frac{1}{aH}} \\ 
	& \approx \frac{H_\ast^2}{8\pi\epsilon_\ast}\LF \frac{k}{k_\ast} \RF^{n_s-1} \frac{1}{2} \LT 2\pm \Delta\Pc_v\LF \frac{k}{k_\ast} \RF \RT \, . 
	\end{aligned}
\label{pw12}
\end{equation}
where 
    \begin{equation}
		\Delta \Pc_v= \LF 2\epsilon+\eta\RF \operatorname{Re}\LT \frac{2}{H_{3/2}^{(1)} \LF \mp k\tau \RF} \frac{\pd H^{(1)}_{\nu_s}\LF \mp k\tau \RF}{\pd\nu_s} \Bigg\vert_{\nu_s=\frac{3}{2}} \RT
		\label{delpP}
	\end{equation} 
From \eqref{power-spectrav} we can deduce that the power spectrum $P_{\zeta_{+}}$ can be mapped to the two-point correlations at $\hat{\textbf{n}}$ and $P_{\zeta_{-}}$ can be mapped to the two-point correlations at $-\hat{\textbf{n}}$ of the CMB sphere. The difference between the two power spectra gives us the non-zero scale-dependent contribution. Production of inflationary correlations in this {direct-sum} Fock space-based quantization can source all kinds of parity-related anomalies, and this can also be interpreted as an indication of spontaneous breaking of time-reversal symmetry in the inflationary background. 
From \eqref{pw12}, we can notice that the two power spectra {corresponding to the two parity conjugate regions of the CMB} differ only by a small scale-dependent correction of the order of the slow-roll parameter. 

\begin{figure}
    \centering
    \includegraphics[width=0.6\linewidth]{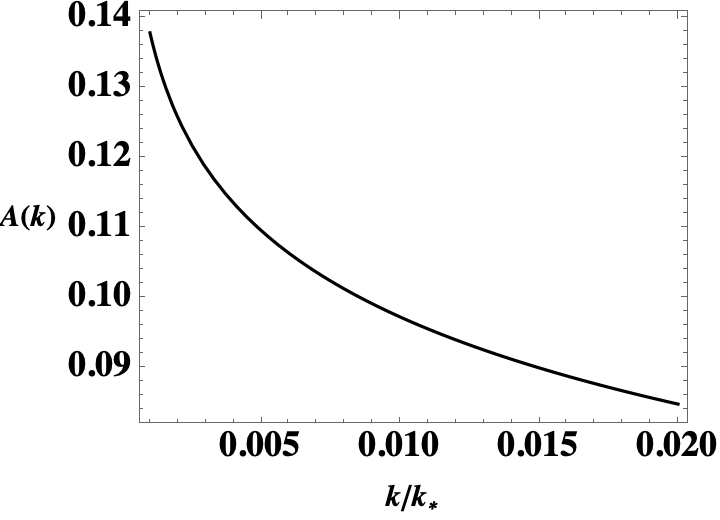}
    \caption{In this figure, we plot the measure of parity asymmetry in the scalar power spectra $A(k)$ in the context of single-field slow-roll inflation for $n_s=0.964$ corresponding to $N=60$ number of e-folds. }
    \label{fig:Ak}
\end{figure}

In addition, through CMB, we can only probe a very limited range of $k$ corresponding to initial $7-8$ e-foldings centered around the pivot scale \cite{Martin:2004um}. 
 At the leading order, the first terms in the two power spectra dominate, which gives us the tilt of the two power spectra {at the parity conjugate points of the CMB} nearly the same in the leading order in slow-roll approximation 
\begin{equation}
	\frac{d\ln P_{\zeta_{+}}}{d\ln k}\approx 	\frac{d\ln P_{\zeta_{-}}}{d\ln k} \approx n_s-1 \approx -2\epsilon-\eta\,.
	\label{tilts2}
\end{equation}
The above result matches the data from the Planck satellite \cite{Gaztanaga:2024vtr}. 

Similarly, we can apply a similar method of quantization for the tensor modes by writing {a tensor fluctuation $h_{ij} =au_{ij}$ as a direct-sum of two {components} in the direct-sum vacuum \eqref{qdSmat} given by}
\begin{equation}
	\begin{aligned}
	\hat{u}_{ij} & = \frac{1}{\sqrt{2}} \hat{u}^{+}_{ij}\LF \tau,\, \textbf{x} \RF \oplus  \frac{1}{\sqrt{2}}\hat{u}^{-}_{ij}\LF -\tau,\,-\textbf{x} \RF \\ & = \frac{1}{\sqrt{2}} \begin{pmatrix}
		 \hat{u}^{+}_{ij}\LF \tau,\, \textbf{x} \RF & 0 \\ 0 &  \hat{u}^{-}_{ij}\LF \tau,\, \textbf{x} \RF
	\end{pmatrix}
\end{aligned}
\end{equation}
The above fields can be written in terms of creation and annihilation operators similar to what we have done for \eqref{vid}. The corresponding mode functions tensor modes can be straightforwardly derived as
\begin{equation}
	\begin{aligned}
		{u}^{\pm}_{ij,\,k} 
		& \approx e_{ij}\sqrt{\frac{1}{2k}} e^{\mp ik\tau}\LF 1 \mp \frac{i}{k\tau} \RF \\ & 
		\pm e_{ij} \epsilon\frac{\sqrt{\pi}}{2\sqrt{k}} \sqrt{\mp k\tau} \frac{\pd H^{(1)}_{\nu_t^{\pm}}\LF \mp k\tau \RF}{\pd \nu_t^{\pm}}\Big\vert_{\nu_t^{\mp}=3/2}
	\end{aligned}
		\label{new-vac1TBD}
\end{equation}
where $e_{ij}$ denotes the polarization tensor. 
As in the case of the scalar power spectra, we obtain two tensor power spectra that describe two-point tensor correlations in the direction $\hat{\textbf{n}}$ and $-\hat{\textbf{n}}$ respectively. The two power spectra of tensor correlations are computed as 
\begin{equation}
	\begin{aligned}
		P_{h_{\pm}} & = \frac{k^3}{2\pi^2}\frac{4}{a^2} \vert {u}^{\pm}_{ij,\,k}\vert^2\Bigg\vert_{\tau = \pm\frac{1}{aH}} \\ 
		& \approx \frac{2H_\ast^2}{\pi}\LF \frac{k}{k_\ast} \RF^{n_t} \frac{1}{2} \LT 2\pm \Delta\Pc_u\LF \frac{k}{k_\ast} \RF \RT \, . 
	\end{aligned}
	\label{pwt12}
\end{equation}
where
\begin{equation}
		\Delta \Pc_u= \LF 2\epsilon\RF \operatorname{Re}\LT \frac{2}{H_{3/2}^{(1)} \LF \mp k\tau \RF} \frac{\pd H^{(1)}_{\nu_s}\LF \mp k\tau \RF}{\pd\nu_s} \Bigg\vert_{\nu_s=\frac{3}{2}} \RT
		\label{delpP1}
	\end{equation}
The tensor power spectrum fractional difference amplitude can be defined as 
\begin{equation}
	T(k) = \frac{P_{h_{+}}-P_{h_{-}}}{4P_h}
	\label{Tko}
\end{equation}
Similar to the derivation of the scalar power spectra tilt \eqref{tilts2}, we also obtain the tilt of the tensor power spectra  in the parity conjugate points of the sky, namely
\begin{equation}
	\frac{d\ln P_{h_{+}}}{d\ln k}\approx 	\frac{d\ln P_{h_{-}}}{d\ln k} \approx n_t \approx -2\epsilon\,.
	\label{tilts3}
\end{equation}

We report our results for the measure of parity asymmetry of primordial power spectra \eqref{Akg} in Fig.~\ref{fig:Ak} and Fig.~\ref{fig:Tk} in the context of single-field $\alpha-$ attractor models represented by the (Starobinsky-like) potential 
\begin{equation}
    V(\phi) = V_0\LF 1-e^{-\sqrt{\frac{2}{3\alpha}}\phi} \RF^2
\end{equation}
where $\alpha$ is a parameter related to the curvature of K\"alher manifold of $\alpha-$ supergravity (SUGRA) theory \cite{Kallosh:2013yoa}. In these models, the slow-roll parameters are 
\begin{equation}
    \epsilon = \frac{3\alpha}{4N^2},\quad \eta = \frac{2}{N}
\end{equation}
where $N$ is the number of e-foldings counted from beginning to ending inflation. In this paper, we consider $N=60$, which gives the value of $n_s= 0.964$, perfectly in line with the Planck data \cite{Planck:2018jri}.

We can notice in Fig.~\ref{fig:Ak} and Fig.~\ref{fig:Tk} that both quantities decrease as we increase the wavenumber.  This is an expected behavior because in the short wavelength modes regime, the curvature of spacetime is locally approaching Minkowski and, therefore, the asymmetry should decrease for the small angular scales or high-$\ell$, a feature compatible with the Planck satellite CMB data \cite{Gaztanaga:2024vtr}. In other words, we see how $\Pc\Tc$ symmetry is broken {(taking into account inflationary background together with quantum fluctuations)} in the { dynamical spacetime as a function of length scales}. Notice that in both Fig.~\ref{fig:Ak} and Fig.~\ref{fig:Tk} we plot the quantities \eqref{Akg} up to the scales $k\lesssim 0.02 k_\ast$ because the power spectra in \eqref{pw12} and \eqref{pwt12} are evaluated at the moment when $k_\ast$ exit the horizon and the only the modes $k\ll k_\ast$ can be assumed to be frozen. The cut-off scale $k_c=0.02k_\ast$ corresponds to the coarse-graining scale of Stochastic inflation deduced from the CMB data analysis in \cite{Gaztanaga:2024vtr}. 

\begin{figure}
    \centering
    \includegraphics[width=0.6\linewidth]{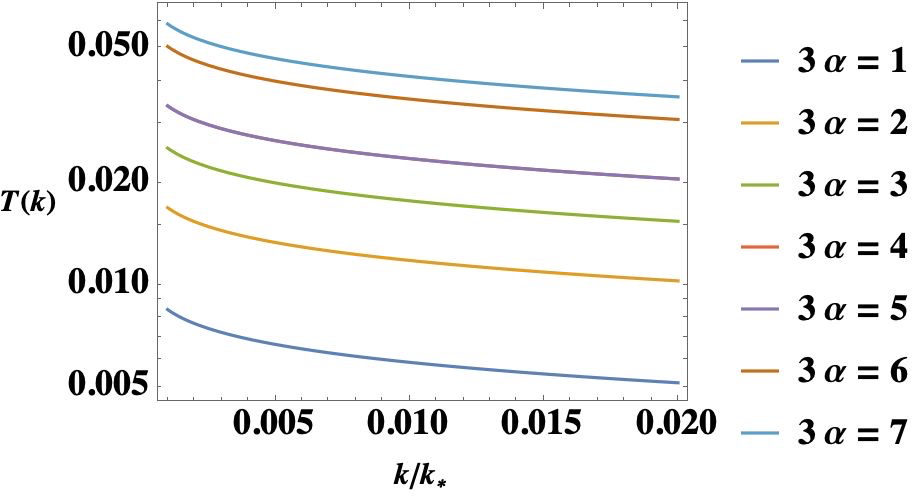}
    \caption{In this figure, we plot the measure of parity asymmetry in the tensor power spectra $T(k)$ in the context of $\alpha-$ attractor model of inflation for $n_s=0.964$ corresponding to $N=60$ number of e-folds. In this plot, we chose preferred values of $\alpha$ from Poincar\'e disc symmetries of SUGRA inflation. The case $\alpha=1$ coincides with the case of Starobinsky inflation. 
    }
    \label{fig:Tk}
\end{figure}

\section{Summary and conclusions}

Inspired by several open questions in the context of QFT in curved spacetime, we propose that the inflationary quantum fluctuations are generated {in a direct-sum quasi-dS vacuum \eqref{qdSmat}} defined by the discrete transformations of spacetime. {Our approach is based on the formulation of quantum theory with two arrows of time using geometric superselection rules. }In the case of curved spacetime, i.e., 
When it involves gravity, one must be careful in understanding spacetime reflection symmetries because gravity introduces dynamics. In our framework, we separate the classical and quantum mechanical notions of time. In inflationary cosmology, the standard procedure is to find a background geometry and quantize gravitational and matter degrees of freedom, respecting the classical notion of time \cite{Martin:2004um}. In our formulation, a quantum fluctuation is the direct-sum of components with opposite time evolutions at parity conjugate points. 
We argue that these quantum fluctuations {in the two spatial sections divided by parity} behave identically in dS spacetime. Still, they are slightly different in the case of inflationary spacetime, which can be interpreted as the spontaneous breaking of {time-reversal symmetry} symmetry. This slight deviation seems to source the parity anomalies in the form of asymmetry in the angular power spectra of even-odd multipoles in the CMB observations. The quantization used to obtain the scalar power spectrum is also applied to the inflationary tensor power spectrum case, and we predict a power asymmetry there as well. Notably, parity asymmetry for scalar and tensor power spectrums is significant only for large angular scales or small wave numbers, and it decreases for small angular scales or large wave numbers.
We quantified all our predictions, which serve as observational tests of our formalism. If future observations targeting the detection of primordial gravitational waves \cite{CMB-S4:2016ple} confirm these results, we will learn significantly about the nature of inflationary quantum fluctuations and QFT in curved spacetime.

\begin{acknowledgments}
KSK acknowledges the support from the Royal Society for the Newton International Fellowship, JSPS, and KAKENHI Grant-in-Aid for Scientific Research No. JP20F20320, and thank Mainz U. for hospitality, where part of the work has been carried out. J. Marto is supported by the grant UIDB/MAT/00212/2020. We want to thank Chris Ripken for the useful discussions and suggestions on the quantization procedure and the initial collaboration on the project.  We thank Gerard 't Hooft for his inspiring talks and discussions about QFT in curved spacetime. We thank Yashar Akrami, Norma G. Sanchez, Masahide Yamaguchi, Alexei A. Starobinsky, Luca Buoninfante, Francesco Di Fillippo, Yasha Neiman, Paolo Gondolo, Martin Reuter, Eiichiro Komatsu, Paulo V. Moniz, Dhiraz Kumar Hazra and L. Sriramkumar for very useful discussions. We also thank Enrique Gazta\~naga for a very useful discussions on observational cosmology. 
\end{acknowledgments}

\appendix



\section{Direct-sum Quantum Mechanics and Quantum Field Theory}
\label{sec:DQM}
  
The concept of describing a single physical world with two distinct arrows of time can also be explored within the framework of the Schr\"{o}dinger equation. This is possible because, in quantum theory, time functions as a parameter rather than an operator, unlike spatial position.  Time is a parameter in quantum theory and not an operator-like spatial position, which dates back to Wigner's insight that time reversal should be represented by an anti-unitary operation.
In general, a positive energy ($\Ec>0$) state in quantum mechanics is always defined with respect to an arrow of time 
\begin{equation}
	\vert \Psi\rangle_t= e^{-i\Ec t_p}\vert \Psi\rangle_0,\quad t_p: -\infty \to \infty
	\label{posten}
\end{equation}
that comes from the form of Schr\"{o}dinger equation 
\begin{equation}
	i\frac{\pd \vert \Psi\rangle}{\pd t_p } = \Ec  \vert \Psi\rangle
	\end{equation}
 We can equivalently write the same \eqref{posten} with the opposite arrow of time, which represents again the same positive energy state due to the sign flip of $i\to -i$ 
 \begin{equation}
 	\vert \Psi\rangle_t= e^{i\Ec t_p}\vert \Psi\rangle_0,\quad t_p: \infty \to -\infty
 	\label{posten2}
 \end{equation}
 The state \eqref{posten2} follows from the Schr\"{o}dinger equation 
 \begin{equation}
 	-i\frac{\pd \vert \Psi\rangle}{\pd t_p } = \Ec  \vert \Psi\rangle
 \end{equation}
 Therefore, even in QM, there are two arrows of time that present a state in the physical world (See also \cite{Donoghue:2019ecz,Donoghue:2020mdd}).
 
Direct-sum quantum mechanics, as developed in \cite{Kumar:2023ctp,Gaztanaga:2024vtr}, accommodates both arrows of time by representing a quantum state as a direct-sum of two distinct components.\footnote{It is essential not to conflate the direct-sum, denoted {$\oplus$}, with standard addition {$+$}. Direct-summing operators result in a block diagonal form, as shown in \eqref{disumS}, and the direct sum of two state vectors forms a higher-dimensional vector, exemplified in \eqref{disumS} \cite{Conway,Harshman,Mazenc:2019ety}}
	\begin{equation}
		\vert \Psi\rangle = \frac{1}{\sqrt{2}} \LF \vert \Psi_+\rangle \oplus \vert \Psi_-\rangle \RF = \frac{1}{\sqrt{2}}\begin{pmatrix}
			\vert \Psi_+\rangle \\ \vert \Psi_-\rangle 
		\end{pmatrix}
		\label{disumS}
	\end{equation}
that evolve with opposite arrows of time at the parity conjugate points of physical space corresponding to geometric superselection sectors\footnote{Superselection sectors are individual Hilbert spaces whose direct sum constitutes the total Hilbert space. In this framework, states within distinct superselection sectors cannot form superpositions with each other \cite{Wick:1952nb,nlab:superselection_theory,Kumar:2023ctp,GKM}. In our approach, these superselection sectors are geometrically defined by parity-conjugate regions of physical space, a departure from conventional interpretations in algebraic QFT  \cite{Wick:1952nb, nlab:superselection_theory,Kumar:2023ctp,GKM}. Accordingly, we term this construction a quantum theory with geometric superselection sectors that laid the foundation for a unitary quantum field theory in curved spacetime \cite{Kumar:2023ctp,Kumar:2023hbj}} of Hilbert space $\Hc = \Hc_+ \oplus \Hc_-$.
The time evolution of the state \eqref{disumS} is governed by what we call the direct-sum Schr\"{o}dinger equation given by
\begin{equation}
i\frac{\pd}{\pd t_p}\begin{pmatrix}
\vert \Psi_+\rangle \\ \vert \Psi_-\rangle 
\end{pmatrix} = \begin{pmatrix}
\hat{H}_+ & 0 \\ 0 & -\hat H_-
\end{pmatrix} \begin{pmatrix}
\vert \Psi_+\rangle \\ \vert \Psi_-\rangle 
\end{pmatrix} 
\end{equation}
where $\hat H= \hat H_+\LF x_+,\, p_+ \RF \oplus  \hat H_-\LF x_-,\, p_-\RF$ is the total (time-independent) Hamiltonian written as direct-sum two components $\hat H_\pm$ which are functions of position and momentum operators $\LF \hat x_\pm,\, \hat p_\pm \RF$. The eigen-values of $\hat x_\pm$ are $x_+ = x\gtrsim 0$ and $x_- = x\lesssim 0$. Note that the operators of one superselection sector do not act on the states of the other by construction \cite{Conway}, for example 
  \begin{equation}
	\begin{aligned}
		\hat{H}\vert \Psi_{+} \rangle & =  \hat{H}_{+}\LF \hat x_+, \hat{p}_+\RF\vert \Psi_{+} \rangle \\
		\hat{H}\vert \Psi_{-} \rangle & =  \hat{H}_{-}\LF \hat x_-, \hat{p}_-\RF\vert \Psi_{-} \rangle
	\end{aligned} 
\end{equation}
The canonical commutation relations in geometric superselection sectors are given by \footnote{Furthermore, we demand the operators corresponding to geometric superselection sectors to commute 
 	\begin{equation}
 		\Big[ \hat x_+,\,\hat x_- \Big]= \Big[ \hat p_+,\,\hat p_- \Big]= \Big[ \hat x_+,\,\hat p_- \Big] = \Big[ \hat p_+,\,\hat x_- \Big] =0\,.
 	\end{equation}
  which is based on the principles of locality and causality.
 }
 \begin{equation}
 	\begin{aligned}
 		[\hat{x}_+,\,\hat{p}_+] & = i\hbar\quad \hat{p}_+= -i\hbar \frac{\pd}{\pd x_+}\quad x_+=x\gtrsim 0 \, \\ 
 		[\hat{x}_-,\,\hat{p}_-] &  = -i \hbar\quad \hat{p}_-= i\hbar \frac{\pd}{\pd x_-}\quad x_-=x\lesssim 0 
 	\end{aligned}
 \end{equation}
 The factor of $\sqrt{2}$ in \eqref{disumS} is a normalization to have the total probabilities add up to 1 as 
 \begin{equation}
 	\int_{-\infty}^\infty dx\langle \Psi \vert \Psi \rangle = \int^{\infty}_0 dx_+\frac{\langle \Psi_+ \vert \Psi_+\rangle }{2}+\int^{0}_{-\infty} dx_{-}\frac{\langle \Psi_{-}\vert \Psi_{-}\rangle }{2}  =1\,,
 \end{equation}
 In direct-sum quantum mechanics, a single positive-energy quantum state is represented as a direct sum of two components. One component evolves forward in time at position $x$ with the time arrow $t_p: -\infty \to \infty$, while the other component evolves backward in time at position $-x$ with the time arrow $t_p: \infty \to -\infty$ 
 \begin{equation}
 	\vert \Psi\rangle_{t_p} = \frac{1}{\sqrt{2}} \begin{pmatrix}
 		\vert \Psi_+\rangle_0 e^{-i\Ec t_p} \\ 
 		\vert \Psi_{-}\rangle_0 e^{i\Ec t_p}
 	\end{pmatrix}
 	\label{staespm}
 \end{equation}
 where $\Ec$ is the energy eigenvalue in the case of the time-independent Hamiltonian. The wavefunction is\footnote{See \cite{Gaztanaga:2024vtr,Kumar:2024quv} for the explicit form of the wave function for the quantum harmonic oscillator.}
 \begin{equation}
 \begin{aligned}
 	\Psi(x,t_p) & = \langle x\vert \Psi\rangle = \begin{pmatrix}
 		\langle x_+\vert & \langle x_-\vert 
 	\end{pmatrix} 
  \begin{pmatrix}
 		\vert \Psi_+\rangle \\ \vert  \Psi_-\rangle 
  \end{pmatrix} 
  \\ & = \frac{1}{2}  \Psi(x_+) e^{-i\Ec t_p} + \frac{1}{2}  \Psi(x_-)e^{i\Ec t_p}
   \end{aligned}
 \end{equation}
 With the above construction, we reconsider the quantum harmonic oscillator through two components of a state, which are "quantum mirror ($\Pc \Tc$)" images of each other (See Fig.~\ref{fig:ho}.)
 \begin{figure}
 	\centering
 	\includegraphics[width=0.6\linewidth]{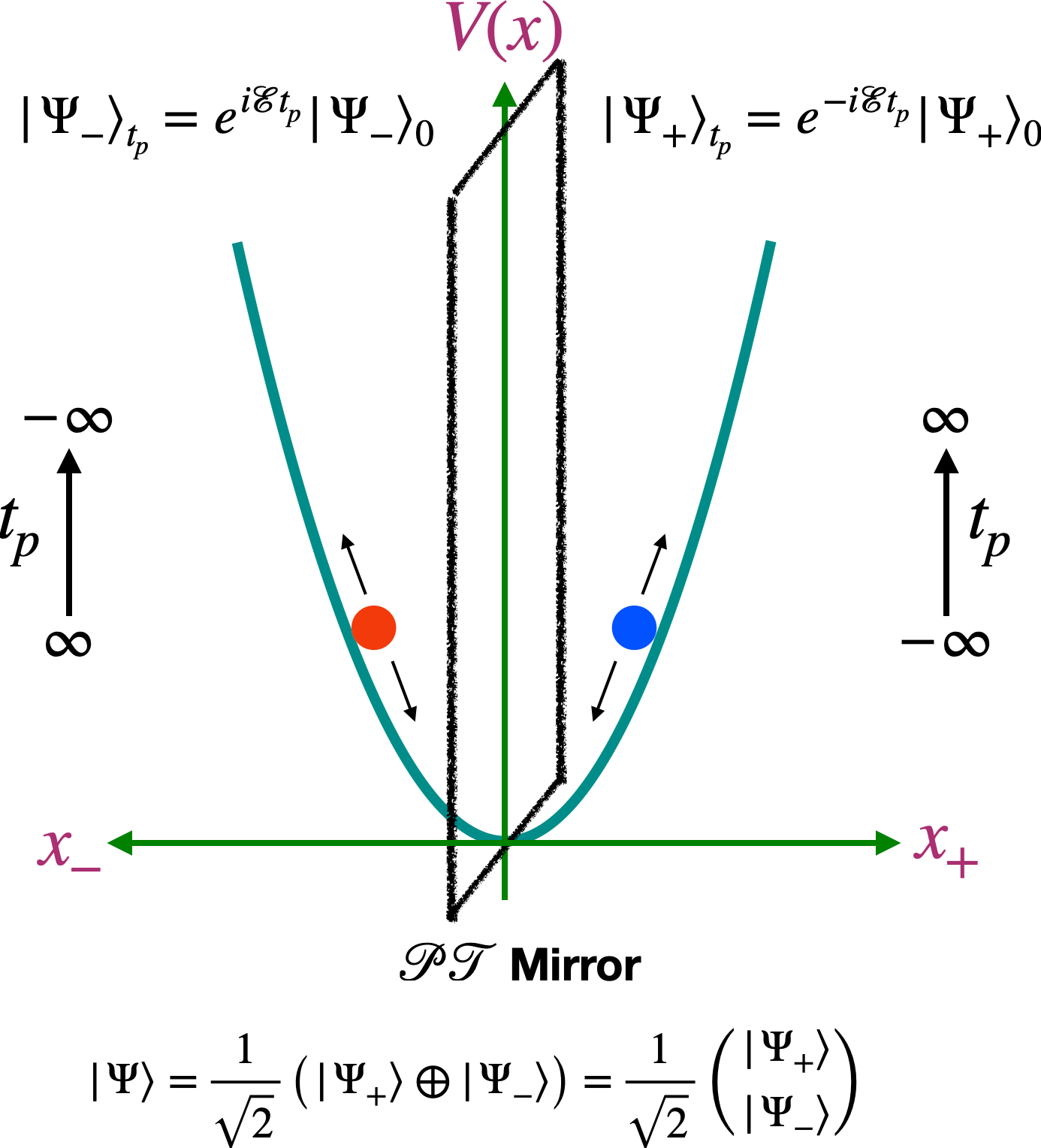}
 	\caption{This figure provides a schematic representation of a quantum harmonic oscillator in the framework of direct-sum quantum mechanics. In this depiction, we observe the two components of a single quantum state of the harmonic oscillator evolving with opposite arrows of time at parity conjugate points in physical space, as governed by the direct-sum Schr\"{o}dinger equation. }
 	\label{fig:ho}
 \end{figure}
 Note that any observable in quantum mechanics is now split into two parts 
  \begin{equation}
 	\langle \Psi \vert \hat \Oc \vert \Psi \rangle = 	\langle \Psi \vert \hat \Oc_+\oplus \hat \Oc_- \vert \Psi \rangle =  \frac{1}{2}	\langle \Psi_+ \vert \hat \Oc_+ \vert \Psi_+  \rangle +\frac{1}{2}	\langle \Psi_- \vert \hat  \Oc_- \vert \Psi_- \rangle \,.
 \end{equation} 

{Direct-sum quantum mechanics can be straightforwardly uplifted to quantum field theory in Minkowski spacetime $ds^2 = -dt_p^2 + d\textbf{x}^2$, which is symmetric under $P: \textbf{x}\to -\textbf{x}$ and $\Tc: t_p\to -t_p$. We propose that the Klein-Gordon (KG) field operator is a direct-sum of the two components 
\begin{equation}
	\begin{aligned}
	\hat{\phi}\LF x \RF & = \frac{1}{\sqrt{2}}  \hat{\phi}_{+}  \LF t_p,\, \textbf{x} \RF \oplus \frac{1}{\sqrt{2}} \hat{\phi}_{-} \LF -t_p,\,-\textbf{x} \RF \\ 
	& = \frac{1}{\sqrt{2}} \begin{pmatrix}
		\hat{\phi}_{+} & 0 \\ 
		0 & 	\hat{\phi}_{-}
	\end{pmatrix}
	\end{aligned}
	\label{disum}
\end{equation}
where $\phi_{\pm}$ are exclusively functions of parity conjugate positions $\pm \textbf{x}$ that are elements of geometric superselection sector Fock spaces ($\Fc_\pm$) expanded in the two commuting sets of creation and annihilation operators as
\begin{equation}
	\begin{aligned}
	 \hat{\phi}_{+}  &  = 	\int \frac{d^3\vec{k}}{\LF 2\pi\RF ^{3/2}	}	\frac{1}{\sqrt{2 \omega_\vec{k}}} \Bigg[\hat a_{+\,\vec{k}}  e^{ik\cdot x}+\hat a^\dagger_{+\,\textbf{k}} e^{-ik\cdot x} \Bigg] \\ 
	 	 \hat{\phi}_{-}   &  = 	\int \frac{d^3\vec{k}}{\LF 2\pi\RF ^{3/2}	}	\frac{1}{\sqrt{2 \omega_\vec{k}}} \Bigg[\hat a_{-\,\vec{k}}  e^{-ik\cdot x}+\hat a^\dagger_{-\,\textbf{k}} e^{ik\cdot x} \Bigg]\,, 
	\end{aligned}
\label{fiedDQFTMin}
\end{equation}
where the operators $a_+,\, a_+^\dagger$ and  $a_{-},\, a_{-}^\dagger$ satisfy the canonical commutation relations, all of their mixed commutators are zero. 
\begin{equation}
	\begin{aligned}
		[\hat{a}_{+\,\textbf{k}},\,\hat{a}_{+\,\textbf{k}^\prime}^\dagger] & = 	[\hat{a}_{-\,\textbf{k}},\,\hat{a}_{-\,\textbf{k}^\prime}^\dagger] = \delta^{(3)}\LF \textbf{k}-\textbf{k}^\prime \RF\\
			[\hat{a}_{+\,\textbf{k}},\,\hat{a}_{-\,\textbf{k}^\prime}] &=	[\hat{a}_{+\,\textbf{k}},\,\hat{a}_{-\,\textbf{k}^\prime}^\dagger] = [\hat{a}_{+\,\textbf{k}}^\dagger,\,\hat{a}_{-\,\textbf{k}^\prime}^\dagger]  =0\,.
		\end{aligned}
	\label{comcan}
\end{equation}
The mixed correlation relations (i.e., the second line in \eqref{comcan}) must be zero to respect locality in the sense that the operators corresponding to forward in time ($t_p: -\infty\to \infty$) at position $\textbf{x}$ should commute with operators corresponding to backward in time ($t_p: \infty\to -\infty$)  at position $-\textbf{x}$. The commutation relations of the field and the corresponding conjugate momenta are given by
\begin{equation}
\begin{aligned}
	[\hat{	\phi}_+\LF t_p, \textbf{x} \RF,\,\pi_+\LF t_p,\,\textbf{x}^\prime \RF] & = i \delta\LF \textbf{x}-\textbf{x}^\prime \RF \\  [\hat{	\phi}_{-}\LF- t_p, -\textbf{x} \RF,\,\pi_{-}\LF -t_p,\,-\textbf{x}^\prime \RF] &= -i \delta\LF \textbf{x}-\textbf{x}^\prime \RF,
	\label{canDQFT}
 \end{aligned}
\end{equation}
where 
\begin{equation}
	\pi_I \LF t_p,\,\textbf{x} \RF = \frac{\pd\Lc_{\rm KG}}{\pd\LF \pd_{t_p}  \phi_+\RF},\quad  	\pi_{-} \LF t_p,\,\textbf{x} \RF = - \frac{\pd\Lc_{\rm KG}}{\pd\LF \pd_{t_p}  \phi_{-}\RF}
\end{equation}
We now define Fock space vacuums as 
\begin{equation}
	\hat{a}_{+\,\textbf{k}}\vert 0\rangle_+ = 0,\quad 	\hat{a}_{-\,\textbf{k}}\vert 0\rangle_{-} = 0 \, ,
\end{equation}
and the total Fock space is direct-sum of $\Fc_\pm$
\begin{equation}
    \Fc = \Fc_+\oplus \Fc_-
\end{equation}
with a direct-sum vacuum 
\begin{equation}
	\vert 0\rangle_T = \vert 0_+\rangle \oplus \vert 0_-\rangle =  \begin{pmatrix}
		\vert 0_+\rangle \\ \vert 0_-\rangle\
	\end{pmatrix}\,.
	\label{tFs}
\end{equation}
Note that both the field operators $\hat \phi_{\pm}$ commute for space-like distance. 

In DQFT, the standard model (SM) degrees of freedom, including particles $\vert SM\rangle$ and antiparticles $\vert \overline{SM}\rangle$ are written as direct-sum of two components in a  direct-sum vacuum \cite{Kumar:2023ctp}:
\begin{equation}
    \vert 0_{SM}\rangle = \begin{pmatrix}
        \vert 0_{SM+}\rangle \\ 
        \vert 0_{SM-}\rangle 
    \end{pmatrix} \quad \vert SM\rangle = \frac{1}{\sqrt{2}}\begin{pmatrix}
        \vert SM_+\rangle \\ 
        \vert SM_-\rangle \end{pmatrix} \quad \vert \overline{SM}\rangle = \frac{1}{\sqrt{2}}\begin{pmatrix}
        \vert \overline{SM}_+\rangle \\ 
        \vert \overline{SM}_-\rangle 
    \end{pmatrix}
\end{equation}
The geometric superselection rule is uniquely defined for all Fock spaces corresponding to the SM degrees of freedom. All Standard Model calculations remain unchanged, as the interaction terms are decomposed into a direct-sum structure in the following way
\begin{equation}
   \Lc_c \sim\Oc_{SM}^3=\begin{pmatrix}
        \Oc_{SM_+}^3 & 0 \\ 
        0 & \Oc_{SM_-}^3
    \end{pmatrix} \quad \Lc_q \sim \Oc_{SM}^4 = \begin{pmatrix}
        \Oc_{SM_+}^4 & 0 \\ 
        0 & \Oc_{SM_-}^4
    \end{pmatrix}
\end{equation}
where $\Oc_{SM}$ is an arbitrary operator involving any SM fields and their derivatives. Furthermore, the $\Cc\Pc\Tc$ invariance holds in both vacuums $\vert 0_\pm\rangle$ \eqref{tFs} and the SM scattering amplitudes remain the same in DQFT \cite{Kumar:2023ctp}.  In a nutshell, the DQFT represents quantum fields and their components as mirror components of parity conjugate regions with opposite arrows of time with the spirit of a new understanding of quantum harmonic oscillator depicted in Fig.~\ref{fig:ho}.   }

\section{Quantum fields in de Sitter spacetime}
\label{app:dS}
Here present the DQFT quantization of a massless scalar field in dS. 
The action of a massless scalar field in dS space is given by
\begin{equation}
	S_{\phi} = -\frac{1}{2}\int d\tau d^3x a^2	\phi\LF \pd_\tau^2+2\Hc \pd_\tau + k^2 \RF \phi\,. 
\end{equation}
The first thing we can notice here is that the above action is invariant under $\Pc\Tc$, which means  $\textbf{x}\to -\textbf{x}$ and 
\begin{equation}\label{timeref}
	t\to -t\quad  \LF-\tau \to \tau\RF \implies H\to -H \quad 
\end{equation}
Rescaling the field $\phi\to a \phi$  gives the following action which leads to Mukhanov-Sasaki equation as
\begin{equation}
	S_\phi = \frac{1}{2} \int d\tau d^3x \Big[ \phi^{\prime 2} - \LF \pd_i\phi \RF^2  +  \frac{2}{\tau^2}  \phi^2 \Big]\,. 
	\label{msaction}
\end{equation}
The above Lagrangian is again symmetric under $\Pc\Tc: \textbf{x}\to -\textbf{x},\, \tau\to -\tau$. When we quantize this scalar field
we propose that the scalar field operator $\hat{\phi}\LF \tau,\, \textbf{x} \RF$ is the direct sum of the pair of field components \cite{Kumar:2023ctp} as
\begin{equation}
	\hat{\phi}\LF \tau,\, \textbf{x} \RF =  \frac{1}{\sqrt{2}}	\hat{\varphi}_{+}\LF \tau,\, \textbf{x} \RF \oplus 	\frac{1}{\sqrt{2}}\hat{\varphi}_{-}\LF -\tau,\, -\textbf{x} \RF \,. 
\end{equation}
which can be expanded in terms of different creation and annihilation operators in the following way
\begin{equation}
	\begin{aligned}
		&	\hat{\varphi}_{+}\LF \tau,\, \textbf{x} \RF   = \frac{1}{\LF 2\pi \RF^{3/2}}\int d\tau d^3k\Bigg[ a_\textbf{k} {\varphi}_{+\,k}\LF \tau \RF e^{-i\textbf{k}\cdot \textbf{x}} + a_\textbf{k}^\dagger {\varphi}^\ast_{+\,k}\LF \tau \RF e^{i\textbf{k}\cdot \textbf{x}} \Bigg]\, 
	\end{aligned}
	\label{vdfield}
\end{equation}
where the creation and annihilation operators $a_\textbf{k},\, a^\dagger_\textbf{k}$ satisfy the canonical commutation relations  and 
\begin{equation}
	a_\textbf{k}\vert 0 \rangle_{dS+} = 0\,. 
 \label{dSp}
\end{equation}
Here the mode function $\varphi_{+\,,k}$ is the solution of Mukhanov-Sasaki equation (that comes from varying \eqref{msaction}) given by 
\begin{equation}
	\varphi_{+,\,k}  = \alpha_{+k} \frac{e^{-ik\tau}}{\sqrt{2k}}\LF 1-\frac{i}{k\tau} \RF +\beta_{+k} \frac{e^{ik\tau}}{\sqrt{2k}}\LF 1+\frac{i}{k\tau} \RF\,, 
	\label{f1}
\end{equation}
where the Bogoliubov coefficients can be fixed as $\LF \alpha_{+k},\, \beta_{+k}\RF = \LF 1,\,0 \RF$, which is compatible with the Wronskian condition $	\varphi_{+,\,k} 	\varphi^{\prime\ast}_{+,\,k}-	\varphi^\ast_{+,\,k}	\varphi^\prime_{+,\,k} = i   $ that corresponds to the canonical commutation relation
\begin{equation} 
	\big[ \hat{\varphi}_{+}\LF \tau,\, \textbf{x} \RF,\, \hat{\pi}_{+}\LF \tau,\, \textbf{x}^\prime \RF\big] = i\delta\LF \textbf{x}-\textbf{x}^\prime \RF. 
\end{equation}
The choice $\LF \alpha_{+k},\, \beta_{+k}\RF = \LF 1,\,0 \RF$ defines the vacuum and $\hat{\varphi}_{+}\LF \tau,\, \textbf{x} \RF \vert 0\rangle $ corresponds to the positive frequency modes of Minkowski vacuum $\vert 0_+\rangle $ \eqref{tFs} in the limit $\tau\to -\infty$. Thus, quantum mechanically $\hat{\varphi}_{+}\LF \tau,\, \textbf{x} \RF \vert 0\rangle $ is the positive energy state that propagates forward in time ($\tau<0,\,H>0$) at \textbf{x}. 
Similarly,  we can expand the second field operator as 
\begin{equation}
	\begin{aligned}
		\hat{\varphi}_{-}\LF -\tau,\, -\textbf{x} \RF   = \frac{1}{\LF 2\pi \RF^{3/2}}\int && d\tau d^3k\Bigg[ b_\textbf{k} \varphi_{-\,k}\LF -\tau \RF e^{i\textbf{k}\cdot \textbf{x}} + \\
		&& b_\textbf{k}^\dagger {\varphi}^\ast_{-\,k}\LF -\tau \RF e^{-i\textbf{k}\cdot \textbf{x}} \Bigg]\, 
	\end{aligned}
	\label{vdfield2}
\end{equation}
where 
\begin{equation}
	\varphi_{-,\,k}  = \alpha_{-k} \frac{e^{ik\tau}}{\sqrt{2k}}\LF 1+\frac{i}{k\tau} \RF +\beta_{-k} \frac{e^{-ik\tau}}{\sqrt{2k}}\LF 1-\frac{i}{k\tau} \RF\,, 
	\label{f2}
\end{equation}
where the Bogoliubov coefficients can be fixed as $\LF \alpha_{-k},\, \beta_{-k}\RF = \LF 1,\,0 \RF$ which is compatible with the Wronskian condition $	\varphi_{-,\,k} 	\varphi^{\prime\ast}_{-,\,k}-	\varphi^\ast_{-,\,k}	\varphi^\prime_{-,\,k} = -i   $ that corresponds to the canonical commutation relation 
\begin{equation} 
	\big[ \hat{\varphi}_{-}\LF -\tau,\, -\textbf{x} \RF,\, \hat{\pi}_{-}\LF -\tau,\, -\textbf{x}^\prime \RF\big] = -i\delta\LF \textbf{x}-\textbf{x}^\prime \RF. 
\end{equation}
which describe the quantum fields that propagate backward in time. 
Here, the second vacuum is defined by 
\begin{equation}
	b_\textbf{k}\vert 0 \rangle_{dS-} = 0\,. 
 \label{dSm}
\end{equation}
Here  $	\hat{\varphi}_{-}\LF -\tau,\, -\textbf{x} \RF\vert 0\rangle_{-}$ corresponds to a positive energy state that evolves backward in time (i.e., $\tau: \infty \to 0,\, H<0$) at -\textbf{x}. 
We demand that these two quantum fluctuations evolve independently. This is manifest by
\begin{equation}
	[	\hat{\varphi}_{+}\LF \tau,\, \textbf{x} \RF,\, 	\hat{\varphi}_{-}\LF -\tau,\, -\textbf{x} \RF   ] =0\,. 
\end{equation}
which implies that the respective creation and annihilation operators commute
\begin{equation}
	\big[a_\textbf{k},\, b_{\textbf{k}^\prime}\big] = 0,\quad \big[a^\dagger_\textbf{k},\, b^\dagger_{\textbf{k}^\prime}\big] = 0 
\end{equation}
The dS vacuum \eqref{dSvacmat} is defined by \eqref{dSp} and \eqref{dSm} which in the short distance limit $\tau\to \mp \infty$ (with $H\to -H$ and $t\to -t$) matches with the positive energy state definitions in Minkowski vacuum \eqref{tFs}. 
Since dS spacetime is perfectly $\Pc\Tc$ symmetric we have that the quantum fields  $	\hat{\varphi}_{+}\LF \tau,\, \textbf{x} \RF\vert 0\rangle_{+}$ and  $	\hat{\varphi}_{-}\LF -\tau,\, -\textbf{x} \RF\vert 0\rangle_{-}$ behave identically, which can be seen from the fact that their equal time correlations are the same
\begin{equation}
	\begin{aligned}
		& \frac{1}{a^2}{}_{dS+}\langle 0\vert \hat{\varphi}_{+}\LF \tau,\, \textbf{x} \RF \hat{\varphi}_{+}\LF \tau,\, \textbf{x}^\prime \RF\vert 0\rangle_{dS+} = \\ & \frac{1}{a^2} {}_{-}\langle 0\vert \hat{\varphi}_{dS-}\LF -\tau,\, -\textbf{x} \RF \hat{\varphi}_{-}\LF -\tau,\, -\textbf{x}^\prime \RF\vert 0\rangle_{dS-} = \frac{H^2}{4\pi^2k^3}\,. 
	\end{aligned}
	\label{eqcorr}
\end{equation}
This makes the correlations of quantum fields, in the parity conjugate regions, identical to the case of dS spacetime.  
We note that the result in \eqref{eqcorr} is similar to the one that was derived in the context of elliptic dS spacetime \cite{SANCHEZ19871111,Schrodinger1956}. However, in this quantization scheme, we do not make any antipodal identification, and therefore, we noticeably deviate conceptually from the proposal of elliptic dS made by Erwin Schr\"{o}dinger \cite{Schrodinger1956}.

\bibliographystyle{utphys}
\bibliography{ssa.bib}

\end{document}